\documentclass{article}
\input epsf.sty
\newcommand{\bright}{\begin{flushright}}
\newcommand{\eright}{\end{flushright}}
\newcommand{\bminip}{\begin{minipage}}
\newcommand{\eminip}{\end{minipage}}
\newcommand{\bcent}{\begin{center}}
\newcommand{\ecent}{\end{center}}

\newcommand{\reflef}{(\ref}

\newcommand{\lsim}{\mbox{\raisebox{-.3em}{$\;\stackrel{<}{\sim}\;$}}}
\newcommand{\beq}{\begin{equation}}
\newcommand{\eeq}{\end{equation}}
\newcommand{\beqa}{\begin{eqnarray}}
\newcommand{\eeqa}{\end{eqnarray}}

\begin{document}
\baselineskip=0.6cm

\bcent
{\Large\bf $\Delta\alpha/\alpha$ from QSO absorption lines driven by an oscillating scalar field}\\[.2em]
Yasunori Fujii$^a$ and Shuntaro mizuno$^b$\\[0.0em]
\baselineskip=0.55cm
$^a$ Advanced Research Institute for Science and Engineering, Waseda University, Tokyo, 169-8555 Japan \\
$^b$ Department of Physics, Waseda University, Tokyo, 169-8555 Japan
\ecent
\mbox{}\\[-3.4em]

\bcent
\bminip{14cm}
\baselineskip=0.5em
{\large\bf Abstract}\\[.6em]
\hspace*{1em}The new result on the QSO absorption lines from the VLT-UVES sample is compared with the past reports on the time-variability of the fine-structure ``constant" derived from the Keck/HIRES observation, on the basis of an oscillatory behavior of the scalar field supposed to be responsible for the cosmological acceleration.
\eminip
\ecent
\baselineskip=0.6cm
\mbox{}\\[-1.6em]
\section{Introduction}

The many-multiplet method, which played a central role in the search for time-variability of the fine-structure ``constant" $\alpha$ from the Keck/HIRES data of QSO absorption lines \cite{mwf,mwf2}, has been applied recently to the VLT-UVES sample, with more restricted selection criteria \cite{chand,srianand}.  The wavelengths of the lines have been determined by two different ways, using either laboratory wavelengths of Mg {\footnotesize II,I} and Si {\footnotesize II} lines with terrestrial isotopic abundances (case 1) or with those of the dominant isotopes (case 2) \cite{chand}.  From case 1, which the authors claim to be more robust, they report no evidence of a changing $\alpha$, in contrast to the previous Keck/HIRES result.  Despite the warning that more careful analysis on the likely systematic errors is necessary before final conclusion is reached \cite{cowie}, it should be worth trying to apply some of the theoretical analyses focusing on different features of the observations accepted at face value for the time being, in order to provide hopefully a guide in entangling complications involved in the phenomenological analyses.

There has been a class of theoretical models in which a cosmological scalar field (dilaton or quintessence) shows itself through a time-dependent fine-structure constant \cite{bj}--\cite{motbar}.  In some of them the scalar field is expected even to be responsible for the cosmological acceleration \cite{bj}--\cite{bento}.

On the cosmological side, the scalar field beyond the linearization regime is supposed generally to be trapped to a potential superimposed on a smooth background, like an exponential potential, thus causing nearly the constant behavior of the sufficient amount of dark energy, the energy density of the scalar field acting as an effective cosmological ``constant." Due to this trapping process, the scalar field $\sigma$ may show a damped-oscillation-like behavior, as a function of the cosmic time $t$, and hence of the fractional look-back time $s = 1-t/t_0$ with $t_0$ the present age of the universe.  We may also assume that the observed coupling strength is proportional to the scalar field, thus expecting $y(s)\equiv \Delta\alpha/\alpha\times 10^5 =K \Delta\sigma(s)$, with $K$ a constant.

Another potentially serious issue comes from the Oklo phenomenon \cite{oklo}--\cite{lamoreaux}.  The detailed analysis of the remnants of the natural reactors supposed to have taken place about 2 billion years ago at Oklo, Gabon, West Africa resulted in either the upper bound $|\Delta\alpha/\alpha|\lsim 1.0\times 10^{-8}$\cite{fifonhom}, or the nonzero values $\Delta\alpha/\alpha \sim (0.9\pm 0.1)\times 10^{-7}$ or $\sim (0.5\pm 0.1)\times 10^{-7}$ for different reasons in Refs. \cite{fifonhom} and \cite{lamoreaux}, respectively.  It has also been argued that it is unlikely that including the effects of the strong interaction enhances the bound more than an order of magnitude \cite{yfptb}.  Comparing these estimates with the values at the level of $10^{-5}$ as expected currently from the QSO observations, we face a difference typically of two orders of magnitude or more.  An oscillatory behavior might provide a possible reconciliation between these two kinds of effects.

As for the suggested virialization process \cite{motbar}, also suggested in this connection, more careful analyses seem yet to be applied in view of our unique features of the scalar field with respect to the structure of matter coupling and the partially finite-range force \cite{cup,separt}.

In view of the uncertainties both on the theoretical and the observational sides, we follow a phenomenological approach to fit the observed $\Delta\alpha/\alpha$ by the assumed simple damped oscillation for $y(s)$, as was attempted in Ref. \cite{plb} applied to the 128 data points from the measurement on the Keck/HIRES QSO absorption lines \cite{mwf}.  We assumed that the Oklo constraint corresponds approximately to a zero of the oscillation at $s_{\rm oklo}= 0.142$, determining the other three parameters by best fitting the QSO data.  We then found that the 3-parameter fit is nearly as good as the weighted-mean-fit as far as the QSO result is concerned \cite{plb}.

We now improve this fit by incorporating a natural zero at $s=0$, arising from the fact that $\Delta\alpha$ is the difference of $\alpha$ from today's value and should vanish at $s=0$ by definition.  This condition is met by offsetting the damped oscillation, as implemented \cite{yfptb} by
%%%%%%%%%%%%%%%%%%%%%%%%%%%%
\beq
y(s) = a\left( e^{bs}\cos \left(v-v_1  \right) -\cos\left( v_1 \right)  \right),
\label{vlt_1}
\eeq
where $v/s=v_1/s_1= v_{\rm oklo}/s_{\rm oklo}=2\pi T^{-1}$  with $v_1$ determined by
%%%%%%%%%%%%%%%%%%%%%%%%%
\beq
v_1 = \tan^{-1}\left( \frac{ e^{-b s_{\rm oklo}} -\cos (v_{\rm oklo}) }{  \sin (v_{\rm oklo})}  \right), 
\label{vlt_2}
\eeq
limiting ourselves at the moment to the behavior that allows many oscillations.   One easily verifies that $y(s)$ vanishes at $s=0$ and $s=s_{\rm oklo}$.  The relaxation time $b^{-1}$ and the period $T$ are both measured in units of $t_0$.

We add that $b\approx 2.5, T\approx 0.22$ are expected from our typical cosmological solutions \cite{cup,prd}, as shown in the beginning of Section 4 of Ref. \cite{plb} and Table 2 of Ref. \cite{yfptb}.  These values will be referred to as ``reference values" in the following, though we do not exclude other solutions.

Equation \reflef{vlt_1}) permits us to compute
%%%%%%%%%%%%%%%%%%%%
\beq
\left( \frac{\dot{\alpha}}{\alpha} \right)_{t_0} = -\frac{1}{t_0}y'(0)\times 10^{-5},\label{vlt_3}
\eeq
which may be different from the averaged rate of change; the weighted-mean divided by the average $\bar{s}$, can be compared directly with the laboratory constraint.

%%%%%%%%%%%%%%% section 2 %%%%%%%%%%%%%%%%
\section{The Keck/HIRES result}

We first apply the fit in terms of \reflef{vlt_1}) to the more recent result of  the 143 data points of Ref. \cite{mwf2}.  The least reduced chi-squared $\chi^2_{\rm rd}= 1.015$ was obtained for $a=0.020, b=5.5,$ and $T=1.352$, as plotted in Fig. 1.  By comparing with their $\chi^2_{\rm rd}= 1.023$ for the simple weighted mean $y=-0.573\pm 0.113$, or the 1-parameter fit in terms of a horizontal straight line, our fit is again nearly as good as theirs, the same feature as was shown in Ref. \cite{plb}.

%%%%%%%%%%%%%%%%% fig 1 %%%%%%%%%%%%%%%%%%
\begin{figure}[th]
\vspace*{-12em}
\hspace*{9em}
%\centerline{\psfig{file= qso03_55_plbin.eps, width=8cm}}
\epsfxsize=8cm
\epsffile{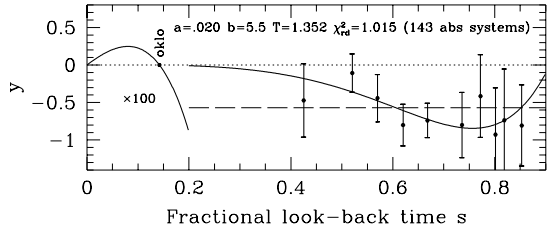}
\vspace*{1pt}
\caption{The best fit to the QSO data in Ref. 2. The portion of the curve for $s<0.2$ is magnified by 100 times.  Only the binned data are shown for simplicity of presentation, but the actual fitting was made for the  entire 143 data points. The dashed line is for the weighted mean, the 1-parameter fit in terms of a horizontal straight line $y=-0.573$, with $\chi^2_{\rm rd}$ = 1.023, with which our 1.015 is nearly comparable.}
\end{figure}

Note that if we were to extend the horizontal straight line, as drawn by a dashed line in Fig. 1, down to $s_{\rm oklo}$, we would have largely missed the point,  contributing an unacceptably large value to the chi-squared because the ``error-bar" to be used in the calculation is much smaller than those for the QSO data, as mentioned before. On the contrary our curve is designed to pass the zero, which should be a good approximation to reality for the same reason.

More detailed comparison does show a decrease of the total amount of $\chi^2$, but it remains too small to pass either of the Akaike or Bayesian information criterion \cite{liddle}.  We emphasize, however, that we have introduced more parameters mainly based on a theoretical ground in connection with the cosmological acceleration, not simply for a better fit.

We find that the preferred value of $T$ is much larger than the reference value, $\sim 0.22$, as explained at the end of the preceding Section.  Such a small value of $T$ is entirely outside the confidence region of 68\%, as shown in Fig. 2.  We may define the mass $m_\sigma = 2\pi/T$.  The  values of $T=1.352$ and $\sim 0.22$ mentioned above correspond to $m_\sigma /H_0=4.75$ and $\sim 32$, respectively.

%%%%%%%%%%%%%%% fig 2 %%%%%%%%%%%%%%%%%%%
%\hspace*{4em}
%\bminip{14cm}
\begin{figure}[th]
\vspace*{-12em}
\hspace*{9em}
%\centerline{\psfig{file=rng103_cmbB.eps,width=8cm}}
\epsfxsize=8cm
\epsffile{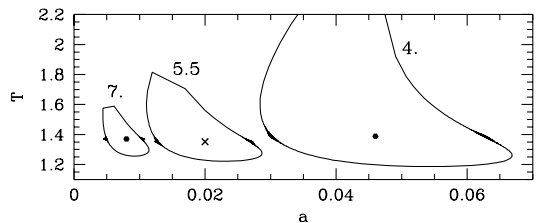}
\vspace*{-1.6em}
\caption{Confidence volume of 68\%, represented by three 2-dimensional contours for $b=4.0, 5.5, 7.0$, respectively, as shown alongside each contour. The cross for $b=5.5$ is for the absolutely minimized $\chi^2_{\rm rd}$, whereas the two dots are only for the 2-dimensional minima for each $b$.}
\end{figure}
%\eminip

To meet the feature of the ``observed" broad distribution, $T=0.87$ and $1.94$ were chosen in Ref. \cite{goldb}, corresponding to $m_\sigma/H_0 = 7.25$ and 3.24, respectively.  They intended, moreover, to respect another constraint from the meteorite dating at $s\approx 0.33$ \cite{olive} by choosing a rather large $b$, if interpreted in terms of \reflef{vlt_1}), resulting in overly pronounced $|y|$ toward the high-$s$ end, thus against the observation.  Nearly the same tendency is found in the analysis of Ref. \cite{gardner} with $m_\sigma/H_0 = 0.24-0.34$, or $T = 26-18$.  In our fit in Fig. 1, we ignored this meteorite constraint according to our own argument \cite{fi,olive2}.  On the other hand, $T\sim 0.5$ has been suggested based on $N=4$ supergravity in Ref. \cite{OB}, though including none of the constraints from Oklo and the meteorite dating.

We find that $(\dot{\alpha}/\alpha)_{t_0} \approx -3.8\times 10^{-17}{\rm y}^{-1}$ turns out to be much smaller than the averaged rate of change, $\sim 0.6\times 10^{-15}{\rm y}^{-1}$.

%%%%%%%%%%%%%%% section 3 %%%%%%%%%%%%%%%%
\section{The VLT-UVES result}

Quite different features appear, on the other hand, to emerge in the most recent report from the VLT-UVES group\cite{chand,srianand}.  The result from 23 data points particularly in case 1 is expressed as a weighted mean $y= -0.06 \pm 0.06$, with $\chi^2_{\rm rd}= 0.95,$ which may be interpreted as no evidence for time-dependence of $\alpha$, contrary to the Keck/HIRES result \cite{mwf,mwf2}.

We attempt the same type of fit in terms of an offset damped oscillator
as in (\ref{vlt_1}).  By transforming the data originally expressed in
terms of redshift, $z$, into a function of $s$ assuming spatially flat
Friedmann cosmology with $t_0= 13.78 {\rm Gy}, h=0.70$, and $\Omega_\Lambda
=0.7$, we found the least $\chi^2_{\rm rd}= 0.53$ for  $a=-0.050, b=3.1,$ and $
T=0.134$, as illustrated in Fig. 3, together with Fig. 4 for the 68\%
confidence region.  Surprisingly, this reduced chi-squared is even
smaller than 0.95 for the weighted mean.  In Fig. 3, we recognize several data points fitted particularly closely by the oscillatory curve, obviously contributing to decrease the chi-squared.

%%%%%%%%%%%%%%%%% fig 3 %%%%%%%%%%%%%%%%%%
\begin{figure}[th]
\vspace*{-12em}
\hspace*{9em}
%\centerline{\psfig{file= vlt_31_mnB.eps, width=8cm}}
\epsfxsize=8cm
\epsffile{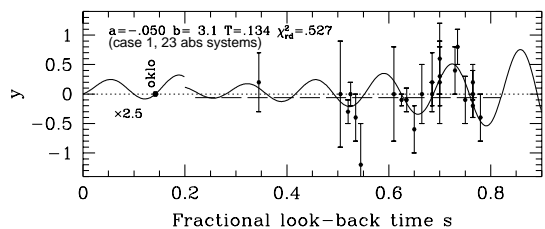}
\vspace*{2pt}
\caption{The best fit to the QSO data for case 1 in Refs. 3,4.  The portion of the curve for $s<0.2$ is magnified by 2.5 times.  The dashed line is for the weighted mean, the 1-parameter fit in terms of a horizontal straight line $y=-0.06$, with $\chi^2_{\rm rd}$ = 0.95, compared with which our 0.53 appears even improved.}
\end{figure}

%%%%%%%%%%%%%%%%% fig 4 %%%%%%%%%%%%%%%%%%
\begin{figure}[th]
\vspace*{-12em}
\hspace*{9em}
%\centerline{\psfig{file= rng3vlt_mnB, width=8cm}}
\epsfxsize=8cm
\epsffile{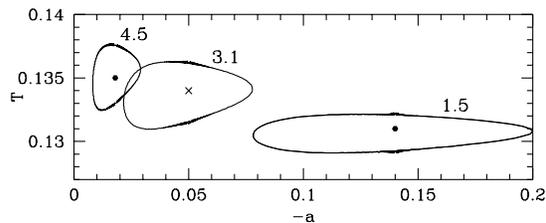}
\vspace*{0pt}
\caption{Confidence volume of 68\% of the fit in Fig. 3, represented in the same manner as in Fig. 2.  Notice $a$ is now negative.}
\end{figure}

The decrease 0.42 in $\chi^2_{\rm rd}$ implies the increase of the $p$-value (in the goodness-of-fit test) by $\sim 45$\% for the small degrees of freedom 23--3=20.  This also results, unlike with the Keck/HIRES data discussed in the preceding section, in the decrease  of the Akaike and Bayesian information criteria (given by $\chi^2 + 2k$ and $\chi^2 + k\ln N$, respectively, with $k$ the number of parameters while $N$ is the number of data points \cite{liddle}) by 6.3 and 4.0, respectively.  These ``improvements" may not be significant enough to select the oscillatory fit unambiguously because the ideal statistical conditions are unlikely met due to possible unknown error sources.  On the other hand, we find that the null result $a=0$ yields $\chi_{\rm rd}^2$ as ``large" as 1.10, entirely outside any of the contours in Fig. 4.  This also implies that our best-fit  function departs from the null result by 2.6 standard deviations.

For these reasons we summarize our analysis by saying that the VLT-UVES data for  case 1 allows a nonzero oscillating $\alpha$, despite a null result \cite{chand,srianand} that is favored by {\it assuming} a uniform time variation, or the 1-parameter fit in terms of a horizontal straight line.  We even argue that choosing a weighted-mean fit is tied with a tacit assumption that the true result is close to the uniform distribution.  On the other hand, the oscillatory fit should be favored as a better starting dependence if there is a theoretical reason to prefer an oscillation, which is motivated by a cosmological acceleration, but tends to be averaged out by an assumed flat distribution.

Furthermore, the favored value of $T$ turns out to be ``small," though even somewhat smaller than the reference value $\sim 0.22$ expected typically from the two-scalar model, showing obviously a pattern significantly different from the ``broad" distribution characterizing the Keck/HIRES result.

Note also that the fitted curve crosses another zero below $s_{\rm oklo}$, and that $(\dot{\alpha}/\alpha)_{t_0} \approx -0.96\times 10^{-15}{\rm y}^{-1}$ is larger in size, contrary to the situation in the Keck/HIRES result, than the averaged rate of change, $0 \lsim \dot{\alpha}/\alpha \lsim 1.2\times 10^{-16}{\rm y}^{-1}$.  We point out that the former estimate is already at the level recently reached by the upper bound $\sim 2.0\times 10^{-15}{\rm y}^{-1}$\ for laboratory measurement \cite{lab}.

We discovered several other fits with $\chi^2_{\rm rd}$ larger than the smallest value 0.527 as shown in Fig. 3, still considerably smaller than 0.95 for the weighted mean. Only two examples are mentioned here; $a=-0.000352, b= 10.0, T=0.156, \chi^2_{\rm rd}= 0.590$ and $a=0.156, b= 1.0, T=0.253, \chi^2_{\rm rd}= 0.777$.  The latter fit is interesting because the obtained $T$ is close to 0.22, the reference value mentioned above.  Without a zero below $s_{\rm oklo}$, unlike in Fig. 3, $(\dot{\alpha}/\alpha)_{t_0}$ is even larger; $\approx +2.7\times 10^{-15}{\rm y}^{-1}$, which might be on the verge of exclusion by the laboratory constraint \cite{lab}.

We also notice that the VLT-UVES result includes the analysis of some sub-samples, giving still smaller chi-squared.  We focused on the 12 absorption systems listed in the first line of Table 5 of Ref. \cite{chand} for ``single + double (case 1)," applying the same fit as for the full sample.  The set of parameters, $a= -0.066, b= 2.55,$ and $T= 0.134$, yields the minimized $\chi^2_{\rm rd}= 0.337$ again smaller than 0.552 for the weighted mean, $y=-0.077\pm 0.101$.  The curve of the fit looks nearly the same as in Fig. 3.  This result seems to demonstrate how robust the presence of an oscillation is.

Having focused on case 1, claimed to be most probably robust, we now apply the same analysis to case 2.  The weighted mean gives $y= -0.36\pm 0.06$ with $\chi^2_{\rm rd}= 1.03$.  This case turns out more like the Keck/HIRES result; a negative $y$ off the zero by 6 sigmas, and apparently broad distribution shown in the damped oscillation fit with $a=0.307, b=0,$ and $T=0.889$, as illustrated in Fig. 5, entailing $\chi^2_{\rm rd}=0.91$ only slightly smaller than that for the weighted mean. Note that $b=0$ does not correspond to the absolute minimum of chi-squared, only at the physical boundary $b\geq 0$.  The fits with $a>0$ result in much larger chi-squared, close to $\chi^2_{\rm rd}=2.52$ for the assumed null result.

%%%%%%%%%%%%%%%% fig 5 %%%%%%%%%%%%%%%%%%
\begin{figure}[th]
\vspace*{-12em}
\hspace*{9em}
%\centerline{\psfig{file= vltc2_0_plB, width=8cm}}
\epsfxsize=8cm
\epsffile{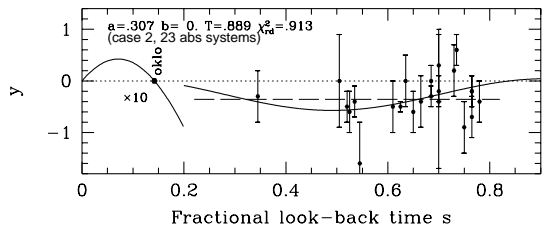}
\vspace*{2pt}
\caption{The best fit to case 2 in Ref. 3, representing the minimum $\chi^2_{\rm rd}$ at the edge of the physically allowed range $b\geq 0$.  The portion for $s<0.2$ is magnified by 10.  The dashed line is for the weighted mean with $y=-0.36$.}
\end{figure}

It is interesting to note that the presence of mass $m_\sigma$ does not prevent $\sigma (t)$ from falling off smoothly, as seen in Fig. 5.8 of Ref. \cite{cup}, Fig. 1 of Ref. \cite{plb} or Fig. 2 of Ref. \cite{prd}.  This represents another mechanism for not disturbing the global structure, somewhat different from the one for a much larger mass of $\sigma$ arising from the self-mass which makes the scalar force of finite-range of the macroscopic order of magnitude \cite{separt}, hence leaving detection of possible WEP violation more remote \cite{cup}.

%%%%%%%%%%%%%%% section 4 %%%%%%%%%%%%%%%%
\section{Discussions}

Our analysis based on the damped-oscillation fit as suggested by a possible connection with the cosmological acceleration showed that the difference between the Keck/HIRES and VLT-UVES (case 1) results goes beyond a question merely of the presence or absence of the time-variability of $\alpha$. It appears as if the oscillation of the scalar field shows quite different periods of time dependence.  Future efforts are expected to separate systematic errors focusing particularly upon a better determination of the period, thus testing the theoretical models favoring various oscillatory behaviors \cite{bj}--\cite{plb},\cite{goldb}--\cite{bento}.

In this respect we also add that there is an exceptional example of non-oscillating behavior in the simplified version of the two-scalar model \cite{cup,prd} obtained by removing the second scalar field \cite{dodel}, though we then lose the aesthetic advantage of the original model, in which no eternal inflation ensues.  The scalar field continues to fall overriding maxima of the sine-Gordon potential included in addition to the background exponential potential until it is finally trapped to one of the minima.  Unlike in the two-scalar model, a mini-inflation (persistenly as one of the repeated occurrences thus lessening the gravity of the coincidence problem) occurs only with a smooth change of the scalar field as illustrated in the lower panel of Fig. 6.  Note that $\Omega_\Lambda$ nevertheless passes through 0.7 as shown in the upper panel, enough to cause an acceleration of the universe.  We failed to detect any oscillation of $\sigma$ generic to the two-scalar model.  This choice, still with the scalar field trapped temporarily, might be favored if future observations disprove oscillatory $\Delta\alpha/\alpha$ and if consistency between QSO and Oklo is achieved by other means.  As shown in Fig. 7, however, the universe today might happen to be entering the final eternal inflation, thus featuring an oscillation, basically as in Refs. \cite{goldb}--\cite{bento}.  Note also that these two different behaviors are discussed in somewhat different context in Ref. \cite{bento}.\\

%%%%%%%%%%%%%%%% fig 6 %%%%%%%%%%%%%%%%%%
\begin{figure}[tbh]
\vspace*{-10em}
\hspace*{9em}
%\centerline{\psfig{file= mzn_1_B1.eps, width=8cm}}
\epsfxsize=8cm
\epsffile{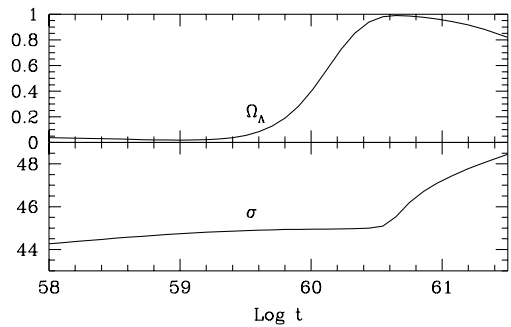}
\vspace*{-2pt}
\caption{A solution in the one-scalar model$^{26}$ with the potential $V(\sigma)= e^{-4\zeta\sigma}[ \Lambda +m^4( 1+\cos(\kappa\sigma))  ]$, with $\zeta=1.5, \Lambda = 1.65 \times 10^{-4}, m=0.42, \kappa =0.351$ in reduced Planckian units, shown only around the present epoch, $t_0\approx 13.8{\rm Gy}$ or ${\rm Log}\: t_0 \approx 60.2$.  The scalar field $\sigma$ varies only smoothly (in the lower panel), whereas $\Omega_\Lambda $ passes through $\sim 0.7$ (in the upper panel), hence causing acceleration of the universe.}
\end{figure}

%%%%%%%%%%%%%%%% fig 7 %%%%%%%%%%%%%%%%%%
\begin{figure}[bth]
\vspace*{-10em}
\hspace*{9em}
%\centerline{\psfig{file= mzn_3_B1.eps, width=8cm}}
\epsfxsize=8cm
\epsffile{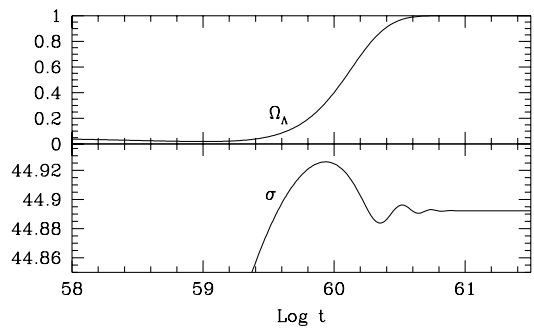}
\vspace*{-2pt}
\caption{A solution for the final eternal inflation for
$\Lambda = 1.54\times 10^{-4}$ in place of the one in Fig. 6 for
repeated mini-inflations.  An oscillatory behavior of $\sigma$ is shown
in the lower panel with the vertical scale magnified by $\sim 45$ times
compared with Fig. 6, nearly the same size of the ``amplitude" as in
Fig. 5.10 of Ref. 7.  For ${\rm Log}\: t < 59.4$, the (invisible)
curve of $\sigma$ is almost identical with the corresponding portion in  Fig. 6.}
\end{figure}

%%%%%%%%%%%%%%%%%%%%%%%%%%%%%%%%%%
\section*{Acknowledgments}

We would like to thank Bruce Bassett, Aldo Fiorenzano, Takashi Ishikawa, Akira Iwamoto, Nobuyuki Kanda, Kei-ichi Maeda, Michael Murphy, Yoshio Oyanagi and Naoshi Sugiyama for many useful discussions.  Special thanks of one of the authors (Y.F.) are also due to Raghunathan Srianand for his generous help in understanding some details of the data.


\begin{thebibliography}{9}
\small

\bibitem{mwf}M.T. Murphy, J.K. Webb, and V.V. Flambaum, 
MNRAS, {\bf 345}, 609 (2003).
\bibitem{mwf2}M.T. Murphy, V.V. Flambaum, J.K. Webb, V.V. Dzuba, J.X. Prochaska, and A.M. Wolfe, Proc. Astrophysics, Clocks and Fundamental Constants, 16--18 June 2003, Bad Honnef, Lect. Notes in Phys. {\bf 648}, 131 (2004), astro-ph/0310318.
\bibitem{chand}H. Chand, R. Srianand, P. Petitjean, and B. Aracil, Astron. Astrophys. {\bf 417}, 853 (2004).
\bibitem{srianand}R. Srianand, H. Chand, P. Petitjean, and B. Aracil, Phys. Rev. Lett. {\bf 92}, 121302 (2004).

\bibitem{cowie}L. Cowie and A. Songaila, Nature {\bf 428} (2004) 132.
\bibitem{bj}Y. Fujii, Int J. Mod. Phys. {\bf D11}, 1137 (2002), First ASTROD School and Symp, 13--23 September 2001, Beijing, astro-ph/0204069; Astrophys. Space Sci, {\bf 283}, 559 (2003), Proc. JENAM 2002, 3--5 September 2002, Porto, Portugal, gr-qc/0212019.
\bibitem{cup}Y. Fujii and K. Maeda, {\sl The scalar-tensor theory of gravitation,} Cambridge University Press, 2003.
\bibitem{plb}Y. Fujii, Phys. Lett. {\bf B573}, 39 (2003).
\bibitem{wett}C. Wetterich, Phys. Lett. {\bf B561}, 10 (2003); D.S. Lee, W. Lee and K.W. Ng, astro-ph/0309316.
\bibitem{goldb}L. Anchordoqui and H. Goldberg, Phys. Rev. {\bf D68}, 083513 (2003).

\bibitem{gardner}C.L. Gardner, Phys. Rev. {\bf D68}, 043513 (2003).
\bibitem{OB}O. Bertolami, R. Lehnert, R. Potting and A. Ribeiro, Phys. Rev. {\bf D69}, 083513 (2004).

\bibitem{bento}M. Bento, O. Bertolami and N. Santos, astro-ph/0402159.

\bibitem{chiba}T. Chiba and K. Khori, Prog. Theor. Phys. {\bf 107}, 631 (2002); D. Parkinson, B.A. Bassett and J.D. Barrow, Phys. Lett. {\bf B578}, 235 (2004); N.J. Nunes and J.E. Lidsey, astro-ph/0310882; P.P. Avelino, C.J.A.P. Martins and J.C.R.E. Oliveira, astro-ph/0402379.   
\bibitem{motbar}D.F. Mota and J.D. Barrow, Phys. Lett. {\bf B581}, 141 (2004) ; MNRAS {\bf 349}, 281 (2004).
\bibitem{oklo}A.I. Shlyakhter, {\it Nature} {\bf 264}, 340 (1976); physics/0307023; T. Damour and F. Dyson, {\it Nucl. Phys.} {\bf B480}, 37 (1996).
\bibitem{fifonhom}Y. Fujii, A. Iwamoto, T. Fukahori, T. Ohnuki, M. Nakagawa, H. Hidaka, Y. Oura, and P. M\"{o}ller, Nucl. Phys. {\bf B573}, 377  (2000); Proc. ND2001, 7--12 October 2001, Tsukuba, Japan, hep-ph/0205206.

\bibitem{yfptb}Y. Fujii, Proc. Astrophysics, Clocks and Fundamental Constants, 16--18 June 2003, Bad Honnef, Lect. Notes in Phys. {\bf 648}, 167 (2004), hep-ph/0311026.
\bibitem{lamoreaux}S.K. Lamoreaux and J.R. Torgerson, Phys. Rev. {\bf D69}, 121701 (2004).
\bibitem{separt}Y. Fujii, Prog. Theor. Phys. {\bf 110}, 433 (2003).
\bibitem{prd}Y. Fujii, Phys. Rev. {\bf D62}, 064004 (2000).
\bibitem{liddle}H. Akaike, IEEE Trans. Auto. Control, {\bf 19}, 716 (1974); G. Schwarz, Ann. Statistics, {\bf 5}, 461 (1978); A. Liddle, astro-ph/0401198.

\bibitem{olive}K. Olive, M. Pospelov, Y.-Z. Qian, A. Coc, M. Cass\'{e}, and E. Vangioni-Flam, Phys. Rev. {\bf D66}, 045022 (2002).
\bibitem{fi}Y. Fujii and A. Iwamoto, Phys. Rev. Lett. {\bf 91}, 261101 (2003).
\bibitem{olive2}K. Olive,  M. Pospelov, Y.-Z. Qian, G. Manh\'{e}s, E. Vangioni-Flam, A. Coc, and M. Cass\'{e}, Phys. Rev. {\bf D69}, 027701 (2004).
\bibitem{lab}M. Fischer {\it et al}, Phys. Rev. Lett. {\bf 92}, 230802 (2004); E. Peik {\it et al}, {\it Phys. Rev. Lett.} {\bf 93}, 170801 (2004).

\bibitem{dodel}S. Dodelson, M. Kaplinghat and E. Stewart, Phys. Rev. Lett. {\bf 85}, 5276 (2000).


\end{thebibliography}
\end{document}